\documentclass[preprint2]{aastex}

\slugcomment{Not to appear in Nonlearned J., 45.} \shorttitle{ New
quasi-periodic oscillations} \shortauthors{Qu et al. }

\begin{document}
\title{Discovery of new quasi-periodic oscillations in the X-ray
transient source V~0332+53}

\author{JinLu, Qu\altaffilmark{1}; Shu, Zhang\altaffilmark{1};
LiMing, Song\altaffilmark{1}; M., Falanga\altaffilmark{2}}

\affil{Laboratory for Particle Astrophysics, Institute of High
Energy Physics, CAS, Yuquan Road 19 (B), 100049 Beijing, P. R.
China} \email{qujl@mail.ihep.ac.cn; szhang@mail.ihep.ac.cn;
songlm@mail.ihep.ac.cn}

\affil{CEA Saclay, DSM/DAPNIA/Service d'Astrophysique (CNRS FRE
2591), F-91191, Gif sur Yvette, France} \email{mfalanga@cea.fr}

\begin{abstract}
We report the discovery of a new quasi-period oscillation (QPO) at
0.22 Hz, centered on the source spin frequency of the high mass
X-ray binary system V~0332+53 when the source was observed during its
November 2004/March 2005 outburst by {\em  RXTE}. Besides this new QPO, 
we also detected  the known 0.05 Hz QPO. Both the 0.22 and 0.05 Hz
QPOs stand out clearly at a mid-flux level of the outburst within
January 15--19 2005, and later at an even lower flux level as the
width of 0.22 Hz QPO drops. No evolution of the centroid frequency
with the flux is seen in either QPO. The rms value below 10 keV is
around 4--6\% for both QPOs  and decreases at higher
energies. We discuss our results in the context of  current QPO models.

\end{abstract}

\section{INTRODUCTION}

The high mass X-ray binary (HMXB)  transient V~0332+53 (X~0331+53)
was discovered during a bright outburst phase in July 1973 with
the  {\em Vela 5B} observatory \citep{terrell84}. After the
discovery, the source was extensively  studied in 
three new  outbursts: in 1983 by \citet{tanaka83} with
{\em EXOSAT}, in 1989 by \citet{makishima90a} with {\em Ginga} and
the last one in 2004/2005 by \citet{zhang05} and \citet{Kreyk05} with {\em
RXTE} and {\em INTEGRAL}. No outburst was detected by BATSE on
board {\em CGRO} during its era between 1991 and 2000  nor by {\em
RXTE} until November 2004 when the source outburst was caught by the
all sky monitor (ASM).

The recurrent HMXB transient system consists of an accreting
neutron star (NS) and an early-type O8--9 Ve optical
counterpart BQ Cam located at a distance of $\sim$ 7 kpc
\citep{bernacca84,honey85,negueruela99}. The pulse period was
first measured at 4.375 s with a 34 day orbit period, and a
moderately eccentric orbit $e \sim 0.3$ \citep{stella85}. The
morphology of the pulse profile changes as a function of energy.
It shows a prominent and broad single-peak at lower-energy and a
secondary peak which becomes more evident above 6 keV
\citep{unger92,Kreyk05}. However, \citet{stella85} and
\citet{zhang05} found that the pulse profile is also dependent on
the source luminosity -- evolving from a  double pulse when the source is
bright to a single pulse when the source flux is lower.  The period
was found to decrease during outburst, with a recently measured
value of $-\dot P\sim 9\times10^{-11}$ s s$^{-1}$  \citep{zhang05}. Possible 
evidence of an accretion disk was found with the detection of a 0.05
Hz QPO \citep{takeshima94}.

Observations \citep{makishima90b,coburn05,soldi05,Kreyk05} 
show that the spectrum is best fitted by an absorbed power-law
with an exponential cut-off and the additional components of
cyclotron absorption lines, located at 25 keV, 50 keV, and 70 keV.
The magnetic field at the NS surface is estimated to be around
2.7$\times 10^{12}$ Gauss.

In this letter we report the discovery of a new QPO based on the
{\em RXTE} observations of V~0332+53 during the 2004/2005
outburst.

\section{OBSERVATIONS and DATA}

\begin{figure}[htb]
\centering 
\includegraphics[width=6 cm, angle= 0]{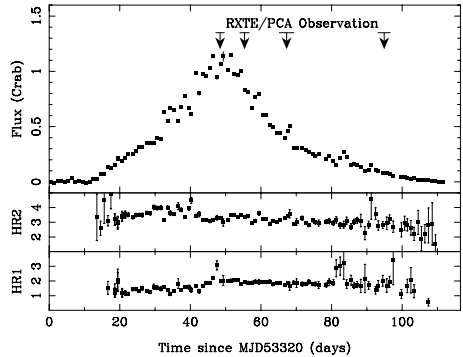}
\caption{ Top panel: one day
averaged {\em RXTE}/ASM light curve in
 the 1.5--12 keV energy band  of  V~0332+53 during the 2004/2005
 outburst. The vertical arrows indicate the times of the {\em
 RXTE}/PCA observations.
Middle panel: hard color (12.0--5.0) keV to  (3.0--5.0) keV. Bottom
panel: soft color (3.0--5.0) keV to (1.5--3.0) keV. }\label{asmlc}
\end{figure}

After the initial follow-up observation by {\em RXTE} starting
November 27, 2004, Target of Opportunity (ToO) observations were
performed between December 28, 2004 and February 15, 2005, the data from
which were made public. The {\it RXTE} observations caught
the source V~0332+53 at the outburst maximum  on December 28--30,
2004 and during the outburst decay phase on January 4--6, 15--19,
and February 12--15, 2005. In Fig.\ref{asmlc} we show the four
time intervals in which the {\em RXTE} observation  was performed.
Hereafter we call the four datasets {\it I}, {\it II}, {\it
  III}  and {\it IV}, respectively.
For our timing analysis, we used these datasets from the
Proportional Counter Array (PCA; 2--60 keV) \citep{jahoda96}. The
PCA data were collected in the binned and event mode, where the
default selection criteria were applied. We used only the PCU2
which was on during the whole observation; the net PCU2
exposure time was 155 ks. The data reduction was performed using
the standard software package FTOOLS version 5.3.1, and the
background contribution was subtracted using the {\it pcabackest} version 4.0.

\begin{figure}[htb]
\centering 
\includegraphics[width=6 cm, angle= 0]{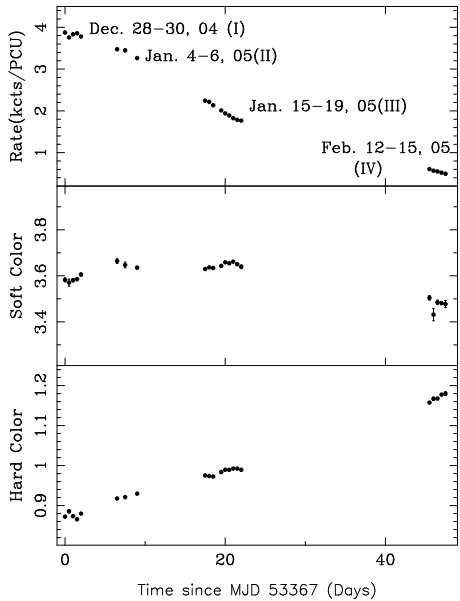}
\caption{The PCA light curve
(2.0-16.5 keV) (top panel) and the time evolutions of soft color
(middle panel) and hard color (bottom panel). Here  MJD~53367
corresponds to December 28, 2004.}\label{pcalc}
\end{figure}

\section{RESULTS}

\subsection{ASM AND PCA LIGHT CURVE}

In Fig.\ref{asmlc} we show the one day averaged ASM light curve of the
entire outburst in the 1.5--12 keV energy band. We converted
the ASM count rate to flux using  1 Crab $\approx$ 75.5 cts/s
\citep{rem04}. The true flux is certain to be substantially higher
since the source is extremely hard, even for X-ray pulsars. The
outburst started on November 23, 2004 (MJD 53332) and reached the
maximum flux level of 1.2 Crab on December 29, 2004 (MJD
53368.6$\pm$0.4). The flux then decayed with an exponential
time-scale,  $F \propto e^{-t/15.9^{d}}$, to quiescence. 
The first evidence of a new X-ray outburst of this source could be
traced back to January 2004 through  observed optical
brightening \citep{gor05}. To analyze the V~0332+53 spectral
variability as function of the outburst time, we defined two
colors. The soft color is the ratio of the 3.0--5.0 keV band to
the 1.5--3.0 keV band and hard color is the ratio of the 5.0--12.0
keV band to the 3.0--5.0 keV band. Colors are built from the one
day averaged ASM light curve. From the color ratios it is evident
that the source underwent a spectral change during the outburst. We
find that the hardness in the soft color increases to a maximum
at the outburst peak, and that the transition in hardness occurs
at an earlier time for the hard color.

In Fig.\ref{pcalc} we show the 0.5 day averaged PCA intensity
evolution of the source in the 2.0--16.5 keV energy band. Using
the 16 s time-resolution Standard 2 mode PCA data we calculated
the  color ratios. The PCA soft color (3.2--5.7 keV/0.7--3.2 keV)
shows a linear trend consistent with the ASM one in the same time
interval. The PCA hard color (9.4--16.5 keV/5.7--9.4 keV) shows a
spectral hardening during the outburst flux decay, compared
to the little change of the spectral shape in the soft color.

\subsection{TIMING ANALYSIS}

We searched for  QPOs in the 2--60 keV energy band, using 
 binned mode and  event mode PCA data. We computed separately
for each of the four datasets a Power Density Spectrum (PDS) in
the frequency range between 1/256 and 8 Hz from Fast Fourier
Transforms. The PDS is normalized using the method described by
\citet{miyamoto91}, and the Poisson white noise is subtracted.
In the resulting PDS two evident QPO signals are present at
$\sim$0.05 and $\sim$0.22 Hz. The  QPO feature
at $0.22$ Hz is newly detected; the low frequency QPO at $0.05$ Hz was
previously observed with {\em Ginga} \citep{takeshima94}. Both
QPOs are detected  in the dataset {\it III} at very high
significance and in the dataset {\it IV} at lower statistical
significance. Outside these two time periods the QPOs are not obvious.
The PDS spectrum also shows  high significant peaks at
the spin frequency, corresponding to a fundamental frequency of 0.228 Hz,
and its higher harmonics. The new QPO is very close to the spin
frequency of the NS. This provides the first detection
of a clear QPO riding on the spin frequency of  a neutron star for a HMXB.

\begin{figure}[htb]
\centering  
\includegraphics[width=6 cm, angle= 0]{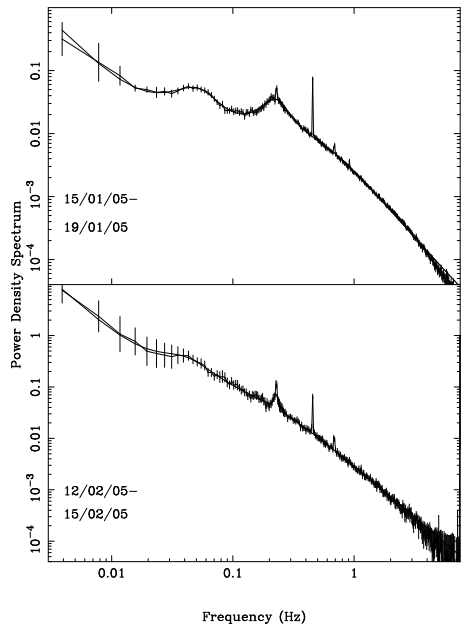} 
\caption{The power density spectra  of V~0332+53 obtained during
the  dataset {\it III} and {\it IV}. The spikes were removed when
the PDS was fitted (see text).}\label{pds}
\end{figure}

The accurate centroid frequency $\nu_0$ and the quality
factor $Q\equiv \nu_0/\delta\nu$ of the QPOs were found 
by fitting the PDS with two Lorentzian functions. 
The broad band-limited noise component
was  fitted using a power-law and a Lorentzian function. The best
fit was also found by removing   the spikes of the spin
frequency and its harmonics (3--5 frequency bins for each spike) from the PDS.
For  dataset {\it III} (January 15--16, 2005), we found the
centroid frequency  for the new QPO at $\nu_0 = 0.223\pm0.002$ Hz
and the quality factor  $Q \approx 2.1$.  The lower frequency
QPO  was at $\nu_0 = 0.049\pm0.007$ Hz with a quality factor of $Q
\approx 1.3$. One month later, using the whole dataset {\it IV} we
found  for the  higher frequency QPO $\nu_0 = 0.227\pm0.002$ Hz
and $Q \approx 12$.

\begin{figure}[htb]
\centering  
\includegraphics[width=6 cm, angle= 0]{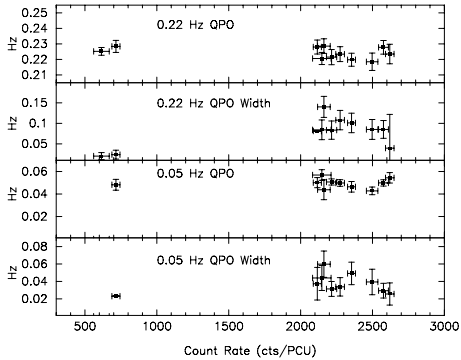} 
\caption{Flux evolutions of the
centroid frequencies  and widths for the 0.22 Hz QPO (first two
panels) and for the 0.05 Hz QPO (lower two panels).}
\label{qpofre}
\end{figure}

The Fourier transform of the time series resulted in the PDS
with best fits  shown in Fig.\ref{pds}. In  Fig.\ref{qpofre} we
show the evolution of the QPO centroid frequencies and widths 
versus the X-ray flux for the datasets {\it  III} and {\it IV}.
We found that the centroid frequencies of both QPOs do not change
significantly with the flux, while the width of the 0.22 Hz QPO
drops with lower flux level.

\begin{figure}[htb]
\centering  
\includegraphics[width=6 cm, angle= 0]{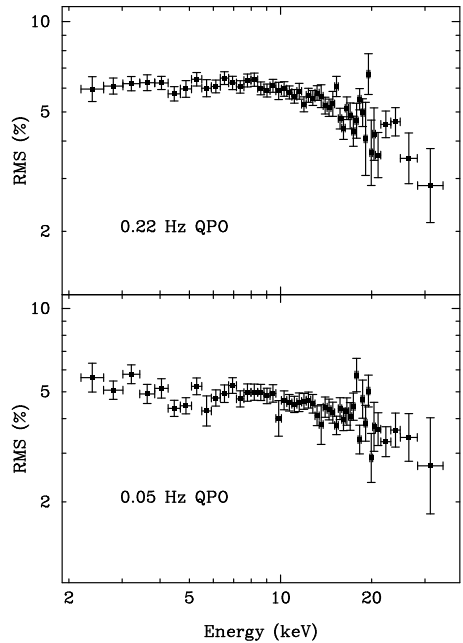} 
\caption{The energy evolutions of
the rms for the 0.22 Hz QPO (upper panel) and for the 0.05 Hz QPO
(lower panel).}\label{rmsqpo}
\end{figure}

To investigate  the QPO energy dependence, we estimated the
fractional root-mean-square (rms) from the PDS. We consider only
the dataset {\it  III} for this analysis, since  the two QPOs
stand out significantly  in this outburst phase. We find for
both QPOs that  the rms is roughly constant below 10 keV and
decreases by a factor of $\sim$2 at higher energies
(Fig.\ref{rmsqpo}). The fact that this trend was not found in the 1989
outburst using {\it  Ginga} data may be due to the relatively  small
effective area of the detector \citep{takeshima94}.

\section{DISCUSSION}

We have discovered a new QPO in V~0332+53  at $\nu_0
= 0.223\pm0.002$ Hz,  close to the spin frequency of the neutron star.
This new QPO was visible in the PDS at the same time as the well known
$\nu_0 = 0.049\pm0.007$ Hz QPO. The
fact that the QPO is found near the spin frequency suggests that it is
produced on or near the surface of the NS. QPOs found in  HMXB pulsars  have
a frequency range of about 10 mHz to 400 mHz \citep[for review
]{finger98,ghosh98} , though none have so far been  detected at the NS
spin frequency.  Broadening features in the PDS at the
spin frequency were detected as well in some other HMXB such as Vela
X-1,  although they were not classified as QPOs \citep{lazzati97}. 
The QPOs found in  HMXBs are considerably lower compared with those found
in  low mass X-ray binary (LMXB) systems which lie in a frequency
range from $\sim6$ to 50 Hz for the low frequency QPO and $\sim300-1200$
Hz for the higher frequency QPO \citep[for review]{klis04}. 
In LMXB, one of the QPOs is interpreted as the NS spin frequency;
this may imply a similar origin for the newly observed QPO 
in the HMXB V~0332+53.

The precise nature of the QPO is still unknown. However, a
series of models have been proposed to explain the QPO features
observed in  compact X-ray binary systems \citep[for
review]{klis04,finger98, ghosh98}. Among  the most frequently used are
the magnetospheric beat frequency model (MBFM) and the Keplerian
frequency model (KFM). In the MBFM, the QPO frequency is
interpreted as the beat frequency between the stellar rotation,
$\nu_s$, and the Keplerian rotation, $\nu_K$, at the inner edge of
the  accretion disk, i.e., $\nu_{QPO}=\nu_K-\nu_s$ \citep{alpar85,
shibazaki87}.  The X-ray flux is proportional to the mass
accretion rate, i.e., $F\propto \dot{M}$.  In the KFM
\citep{klis87}, the QPOs arise from the modulation of the X-rays by
inhomogeneities in the inner disk at the Keplerian frequency,
i.e., $\nu_{QPO}=\nu_K$, and the X-ray flux is $F\propto\nu_K^{7/3}$.

Both models predict a systematic change of the QPO frequency with
the mass-accretion rate or with the flux. However, as shown in
Fig.\ref{qpofre}, such a feature is not observed in V~0332+53.
From  dataset {\it III} to  {\it IV} the   X-ray flux changed by a
factor of $\sim4.5$, but the centroid frequencies of both QPOs remain constant
within the error bars. This suggests that neither the MBFM nor the
KFM can explain the QPO phenomena as observed in V~0332+53. Similar
conclusions have been drawn by different authors. For example, the
expected  luminosity related to the QPO frequency does not always
agree with the observed luminosity
\citep{finger98}. Other observations have not been supported by the
MBFM model, e.g., as found by Takeshima et al. (1994) through the
comparison  of the estimated Alfv\'en radius from  accretion disk
theory  (Ghosh \& Lamb 1979a,b).

One alternative model to explaining such correlations was given by
\citet{lazzati97} and \citet{burderi07}. They argued that the
accretion flow is inhomogeneous near the surface of the neutron
star, and  random shots are characterized by an arbitrary
degree of modulation. The combination of rotation and radiative
transfer effects should therefore produce a periodic modulation of
the shots similar to that of any continuum X-ray emission from 
the polar caps. They proposed that a coupling between the periodic
and red-noise variability should be frequently present in X-ray
pulsars. The discovery of a new QPO with the frequency of the spinning
neutron star provides an additional example which could be
understood within the scenario of this model.

\acknowledgments J.L. Qu thanks F. J. Lu for helpful discussions.
This work was subsidized by the Special Funds for Major State
Basic Research Projects and by the National Natural Science
Foundation of China. This research has made use of data obtained
through the High Energy Astrophysics Science Archive Research
Center Online Service, provided by the NASA/Goddard Space Flight
Center.

\clearpage

\clearpage


\end{document}